\documentclass[%
 reprint, 
 amsmath,amssymb,
 aps,
 nofootinbib
 prb,
 superscriptaddress
 ]{revtex4-1}
\usepackage{scrextend}
\usepackage{graphicx}
\usepackage{dcolumn}
\usepackage{bm}
\usepackage{tabularx}
\usepackage[utf8]{inputenc}
\usepackage[english]{babel}
\usepackage[bottom]{footmisc}
\usepackage{color}

\usepackage{amsmath}
\usepackage{soul,xcolor}
\setstcolor{red}
\usepackage{hyperref}
\usepackage{booktabs}
\usepackage{epstopdf}

\DeclareMathAlphabet{\pazocal}{OMS}{zplm}{m}{n}

\begin{document}

\preprint{APS/123-QED}

\title{Promising High Temperature Thermoelectric Performance of Alkali Metal-based Zintl phases
X$_2$AgY (X = Na, K; Y = Sb, Bi): Insights from First-Principles Studies}

\author{Mohd Zeeshan}
\affiliation{Department of Physics, Indian Institute of Technology,
             Hauz Khas, New Delhi 110016, India}

\author{Indranil Mal}
\affiliation{Department of Physics, Indian Institute of Technology,
             Hauz Khas, New Delhi 110016, India}

\author{B. K. Mani}
\email{bkmani@physics.iitd.ac.in}
\affiliation{Department of Physics, Indian Institute of Technology,
Hauz Khas, New Delhi 110016, India}

\date{\today}
             
\begin{abstract} 

In the quest for novel thermoelectric materials to harvest waste environmental
heat, we investigate alkali metal-based Zintl phases X$_2$AgY (X = Na, K, and
Y = Sb, Bi) utilizing first-principles methods. We obtain significantly low
lattice thermal conductivity values ranging 0.9-0.5 W m$^{-1}$ K$^{-1}$ at 300~K,
challenging established thermoelectric materials such as SnSe, PbTe, Bi$_2$Te$_3$
as well as other Zintl phases. We trace such astonishingly low values to
lattice anharmonicity, large phonon scattering phase space, low phonon velocities,
and lifetimes. In K-based materials, the low phonon velocities are further linked
to flattened phonon modes arising from the gap in the optical spectrum. Furthermore, the
existence of bonding heterogeneity could hamper heat conduction in these materials.
In addition, an avoided crossing in the phonon dispersions suggesting rattling behavior, observed in all materials
except Na$_2$AgSb, suppresses the dispersion of acoustic modes, further reducing the phonon
velocities. When combined with electrical transport calculations, the materials exhibit high
figure of merit values at 700~K, i.e., $ZT\sim2.1$ for Na$_2$AgSb, $1.7$ for Na$_2$AgBi,
$0.9$ for K$_2$AgSb, and $1.0$ for K$_2$AgBi. Our predicted $ZT$ values are competitive
with state-of-the-art thermoelectric materials such as Mg$_3$Sb$_2$, ZrCoBi, PbTe, SnSe,
and as well as with contemporary Zintl phases. Our findings underscore the potential of light
alkali metal atoms combined with Ag-Bi/Sb type frameworks to achieve superior thermoelectric
performance, paving the way for material design for specific operating conditions.

\end{abstract}

\maketitle

\section{Introduction}

World's primary energy demand is still predominantly met by the
consumption of fossils \cite{Zoui20}. However, fossils are nonrenewable and
will not last forever \cite{Hamid14}. The current energy consumption rate and
increasing population demand will likely deplete the resources faster \cite{Arut17}.
On the other hand, the bleak efficiency of the
energy produced is also concerning \cite{Forman16}. Almost 60\% of the energy
produced is lost to the environment in the form of heat \cite{Zhang10}. This
alarming rate of heat rejection to the environment could lead to
climate changes. It is also well known that the combustion
of fossils contributes to greenhouse gases, ultimately
facilitating global warming \cite{Jackson22}.
Thus, there is a pressing need and high time to look beyond
conventional power generation methods. Various clean
renewable alternatives have been devised, such as solar
cells \cite{Liu23}, hydrogen storage materials \cite{Zuttel03},
and many others \cite{Ang22}.
However, these technologies are yet to find a
firm footing as alternative sources of power generation.
Researchers around the globe have been actively engaged
in overcoming the hurdles associated with sustainable
power generation.

In the same endeavor, thermoelectricity has emerged as
a promising alternative for power generation \cite{Singh24}.
The technology can harvest the environment's enormous waste
heat into electricity \cite{Wehbi22}.
This will eventually ease the burden on the existing
resources by improving the efficiency of energy
usage. Further, if scaled well, thermoelectricity
could be a plausible alternative source of power generation.
Despite the stated benefits, the commercialization
of these materials is hindered by their low efficiency \cite{Garmroudi21}.
The efficiency of thermoelectric material is based
on the temperature gradient and a dimensionless quantity
called the figure of merit, which is expressed as $ZT = S^2 \sigma
T$/$(\kappa_e + \kappa_L)$, where $S$, $\sigma$, $T$,
and $\kappa$ are the Seebeck coefficient, electrical conductivity,
temperature, and thermal conductivity including electronic
and lattice thermal contributions \cite{Karmakar20, Maassen21, Berry17}.
The electronic part ($S$, $\sigma$,
and $\kappa_e$) is heavily interconnected and poses a challenge
for simultaneous optimization, resulting in lower thermoelectric
efficiency \cite{Boona17, Lee17, Bell08}.

The two distinct strategies to improve the figure of merit are
improving the numerator part, $S^2\sigma$, also called as
power factor, and lowering the $\kappa_L$ \cite{Zeeshan18}. The power factor
has been successfully improved using band engineering \cite{Xi16},
optimizing carrier concentration \cite{Wei15}, and resonant
doping \cite{Xin23}. On the other hand, the $\kappa_L$ is significantly
lowered by adopting nanostructuring \cite{Snyder08}, introducing defects
through vacancies or alloying \cite{Chen13},  and including grain boundaries
\cite{Dehkordi15}.
However, the two approaches often interfere, ultimately deteriorating the overall
figure of merit \cite{Chandan24}.
Interestingly, many systems possess inherently low $\kappa_L$
\cite{Dutta19, Jana16, Chen21, Qiu16, Li20}.
Such systems derive low $\kappa_L$ through intrinsic rattlers,
hierarchical bonding, complexed crystal structures, anharmonicity
in lattice vibrations, low sound velocities, and high atomic
displacements. The inherently low $\kappa_L$, which is nearly
independent of the electronic part, allows to independently tune
the electronic part, i.e., the basis of Slack's phonon-glass
electron-crystal (PGEC) concept. According to the PGEC concept \cite{Slack95},
a good thermoelectric material should possess glass like
$\kappa_L$ and crystal like electrical conductivity.

Zintl phases, in principle, have been great contenders
in recent times to follow the PGEC concept to a certain extent
\cite{Wang20, Balvanz20, Chen19}.
Numerous Zintl phases are reported to exhibit good thermoelectric
properties based on inherently low $\kappa_L$ and reasonable power
factor. For instance, Yb$_{14}$MnSb$_{11}$ \cite{Roude14} ($ZT\sim$ 1 at 1200~K),
Ca$_5$Al$_2$Sb$_6$ \cite{Toberer10} ($ZT>$ 0.6 at 1000~K),
Sr$_3$AlSb$_3$ \cite{Zevalkink13} ($ZT$ $\sim$ 0.3 at 600~K), and BaCuSb \cite{Zheng22}
($ZT \sim$ 0.48 at 1010~K) derive good
figure of merit values based on their inherently low $\kappa_L$.
It is important to note that Zintl phases cover extensively a vast
range of systems. Compounds fulfilling the criterion of
electropositive ions donating electrons to the otherwise anionic
framework lie under the umbrella of Zintl phases \cite{Kauzlarich23}. However, not all
Zintl phases could be potential thermoelectric candidates. Since the
Zintl family is large and continuously growing, identifying the
potential thermoelectric Zintl phases could significantly contribute
to the ongoing research.

First-principles simulations can greatly facilitate the cause and
provide a roadmap for experimental investigations. For instance,
in Ref.~\cite{Pal18}, authors investigated that bonding hierarchy led to high
thermoelectric performance in Zintl phase BaAu$_2$P$_4$. In another
study, Wang \textit{et al}. \cite{Wang23} investigated nine Zintl phases and found
a high figure of merit for SrAgSb ($ZT \sim$ 0.94 at 400~K), based on
enhanced phonon scattering due to softening of acoustic phonons. Very
recently, Wei \textit{et al}. \cite{Wei23} reported a high figure of merit for Zintl
compound KSrBi ($ZT \sim$ 2.8 at 800~K). This was attributed to the low
$\kappa_L$ arising from rattling vibrations and occupied antibonding
states. Motivated by such recent findings, we decided to
computationally explore unreported Zintl phases for thermoelectrics.

In this work, we use first-principles based methods to systematically
investigate the electronic and thermal transport properties of the alkali
metal-based family of Zintl phases X$_2$AgY (X = Na, K, and Y = Sb, Bi).
We report significantly low lattice thermal conductivity values competing
well with established thermoelectric materials such as SnSe, PbTe, and
Bi$_2$Te$_3$. To account for such low values, we investigate different
aspects of thermal transport, such as phonon scattering phase space,
velocities, lifetimes, and lattice anharmonicity. In all systems except
Na$_2$AgSb, we find an avoided crossing characteristic of rattling behavior
associated with suppressing the dispersion of acoustic modes. Such a
phenomenon leads to reduced slopes and low phonon velocities. Particularly
in K-based materials, we find gap in optical modes, which often results in
the flattening of phonon modes. Furthermore, we obtain bonding heterogeneity,
which is disruptive to thermal conductivity. Combining such compelling
thermal transport results with electrical ones, we further report tremendous
figure of merit values for Na$_2$AgSb and Na$_2$AgBi. Our study signifies the
microscopic origin of low lattice thermal conductivity values and regulation
of carrier engineering in optimizing thermoelectric materials.

The paper is arranged as follows: In Sec.~II, we briefly provide
the computational methods used for performing our simulations.
In Sec.~III, we elucidate our results on structural optimization,
thermal transport, and electrical transport properties. In Sec.~IV,
we discuss the aspects pertaining to the experimental realization of proposed
properties and also highlight the prospects of the work. To
conclude, a summary of the work is presented in Sec.~V of the
paper.

\section{Computational Methods}

First-principles simulations based on density functional theory
(DFT) are performed using Vienna \textit{ab initio} simulation
package (VASP) \cite{Kresse96, KressePRB}. The interactions between
core and valence electrons are accounted using projector-augmented
wave pseudopotentials. The exchange-correlations among electrons are
treated using the generalized gradient approximation (GGA) based
Perdew-Burke-Ernzerhof (PBE) pseudopotential \cite{Perdew96}. The cutoff
energy of 500 eV is used for the plane-waves for all four systems.
The Monkhorst-Pack scheme is used for Brillouin zone integration.
The ground state structural parameters are obtained using a \textit{k}
grid of 11 $\times$ 11 $\times$ 11, whereas a denser grid of 21
$\times$ 21 $\times$ 21 is used for self-consistent energy
calculations, ensuring sufficient accuracy for transport
properties. The total energy and forces convergence criteria
are set as 10$^{-8}$ eV and 10$^{-7}$ eV/Å, respectively.

The electronic structure is calculated using the modified
Becke-Johnson (mBJ) potential \cite{Tran09}. In addition,
relativistic effects are included to account for the contribution
of heavy atoms in the proposed systems. We employed
TB-LMTO-ASA code \cite{Jepsen99} to investigate the crystal orbital
Hamilton population (COHP) and bonding characteristics.
Based on the obtained
electronic structure, we have calculated the electrical transport properties using the
AMSET code \cite{Ganose21}. We have considered three scattering mechanisms as
implemented in the code, viz. acoustic deformation potential,
ionized impurity, and polar optical phonon. The
net scattering rates are evaluated based on Matthiessen’s rule \cite{Matt64}
\begin{equation}
 \frac{1}{\tau_\mathrm{e}} = \frac{1}{\tau_\mathrm{{ADP}}} + \frac{1}{\tau_\mathrm{{IMP}}} + \frac{1}{\tau_\mathrm{{POP}}}
\end{equation}
The components of scattering rates are obtained as per the
Fermi's golden rule \cite{Dirac27}
\begin{equation}
\tilde{\tau}^{-1}_{n \mathbf{k} \rightarrow m \mathbf{k} + q} = 
\frac{2 \pi}{\hbar} \arrowvert g_{nm} (\mathbf{k},\mathbf{q}) 
\arrowvert^2 \delta (\epsilon_{n \mathbf{k}} - \epsilon_{m \mathbf{k} + q}) 
\end{equation}
where $n$\textbf{k} represent the initial wave vector, $m$\textbf{k} +
\textbf{q} is the final waver vector, $\hbar$ is the reduced
Planck’s constant, $\delta$ is the Dirac delta function, $\epsilon$
is the electron energy, and $g$ is the electron-phonon coupling
matrix element. The input parameters required for calculated scattering
mechanisms, such as elastic and dielectric constants, deformation potential, and polar
optical phonon frequency, are extracted using first-principles simulations,
density functional perturbation theory, and finite differences method.
The electrical transport coefficients are well tested in terms of the interpolation
factor, which converges well with a value of 5.

A supercell of 2 $\times$ 2 $\times$ 2 size containing
128 atoms is used for generating the force constants
by finite displacement method along with $\Gamma$-centered
$k$ mesh of 2 $\times$ 2 $\times$ 2. The atomic displacements
of 0.01 Å and 0.03 Å are used for second- and third-order force
constants, respectively. The phonon dispersion curves are obtained
using the Phonopy \cite{Togo15} code by solving the expression
\begin{equation}
\sum_{\beta\tau'} D^{\alpha\beta}_{\tau \tau'}
(\mathbf{q}) \gamma^{\beta\tau'}_{\mathbf{q}j} = 
\omega^2_{\mathbf{q}j}\gamma^{\alpha\tau}_{\mathbf{q}j}. 
\end{equation}
where the indices $\tau, \tau'$ represent the atoms, $\alpha, \beta$
are the Cartesian coordinates, ${\mathbf{q}}$ is a wave vector,
$j$ is a band index, $D(\mathbf{q})$ is the dynamical matrix,
$\omega$ is the corresponding phonon frequency, and $\gamma$
is the polarization vector.
To compute the Grüneisen parameter, the phonon
calculations are performed by scaling the unit-cell volume by $\pm$2\%.

The lattice thermal conductivity is determined by solving the
Boltzmann transport equation for phonons within the single-mode
relaxation time approximation (RTA) using the Phono3py \cite{TogoPRB}
code. The intrinsic values are obtained through phonon lifetimes
$\tau_\lambda$, group velocities $\mathbf{v_\lambda}$, and
mode heat capacities $C_\lambda$, according to the
relation
\begin{equation}
\kappa_L = \frac{1}{NV} \sum_\lambda C_\lambda \mathbf {v}_\lambda 
\otimes \mathbf {v}_\lambda \tau_\lambda
\label{eqn}
\end{equation}
where $N$ represents the number of unit-cells of the crystal and $V$ is
the volume of the unit cell. The lattice thermal conductivity is well
converged in terms of cutoff pair distance (5~\AA) and \textit{q}-grid
(17 $\times$ 17 $\times$ 17 $q$). The lifetime of phonons
is determined from third-order force constants, which accounts for phonon-phonon
scattering. To account for three-phonon scattering processes, we calculated
phase space using ShengBTE \cite{Sheng14} code.

\begin{figure*}
\centering\includegraphics[scale=0.27]{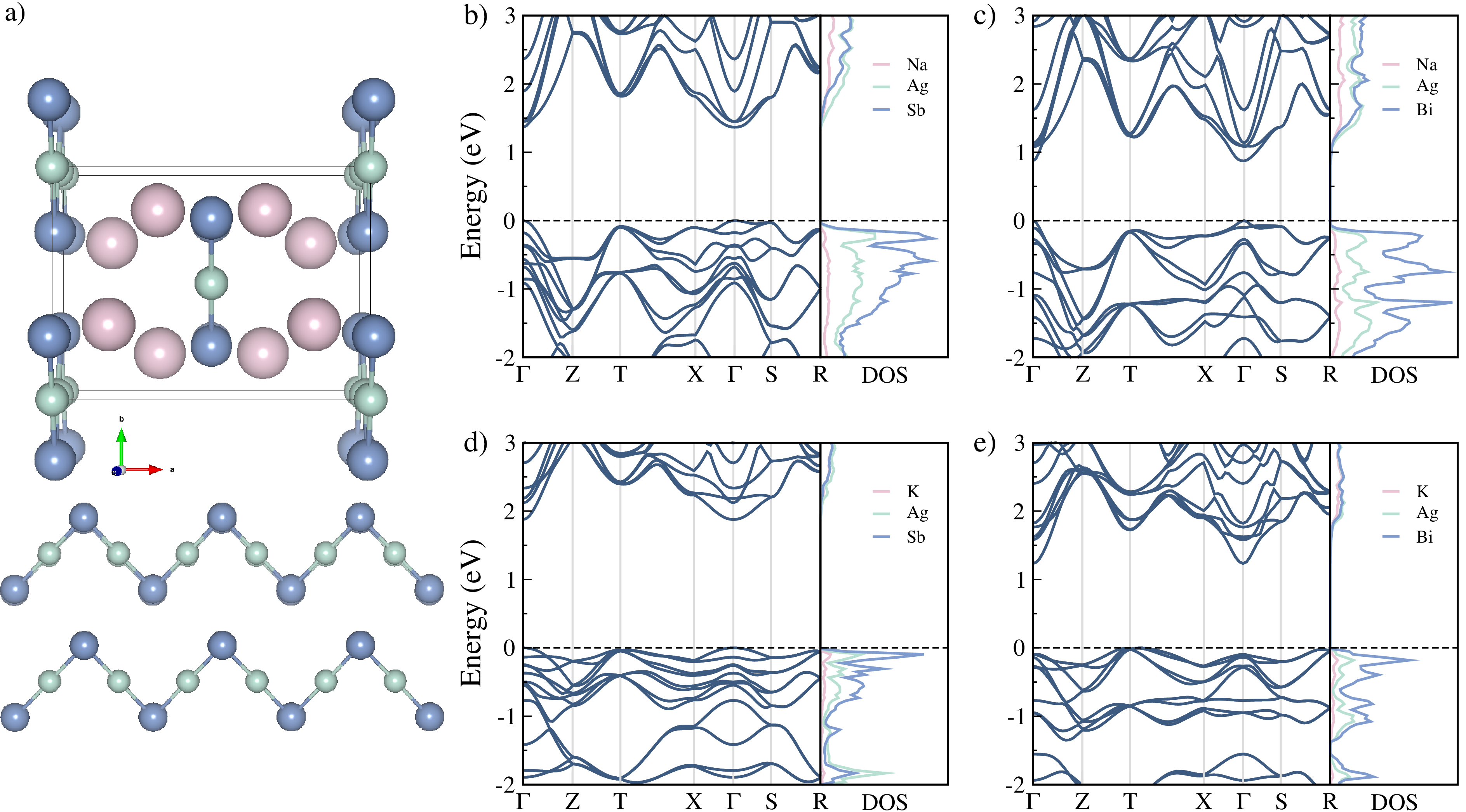}
\caption{(a) Crystal structure of X$_2$AgY (X = Na/K, Y = Sb/Bi) in
orthorhombic $Cmcm$ symmetry and the zigzag alignment of Ag-Sb/Bi,
(b), (c), (d), and (e) are electronic structures of Na$_2$AgSb, 
Na$_2$AgBi, K$_2$AgSb, and K$_2$AgBi, respectively. The density
of states is shown in arbitrary units.} 
\label{crystal}
\end{figure*}

\section{Results}

\subsection{Structural Optimization}

\begin{figure}
\centering\includegraphics[scale=0.35]{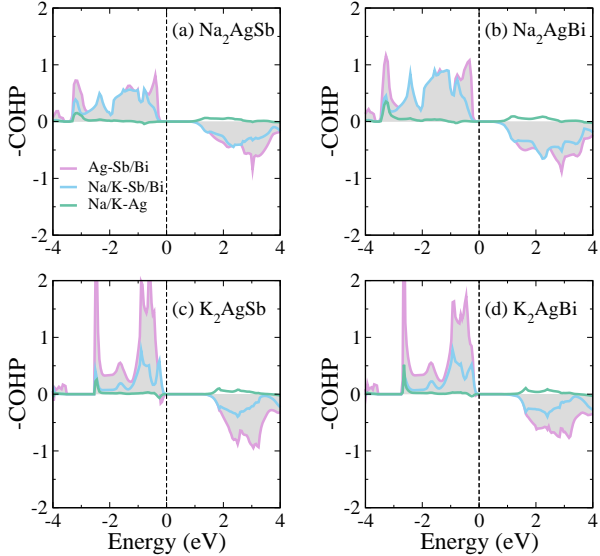}
\caption{Crystal orbital Hamiltonian population of (a) Na$_2$AgSb,
(b) Na$_2$AgBi, (c) K$_2$AgSb, and K$_2$AgBi.}
\label{cohp}
\end{figure}

\begin{figure*}
\centering\includegraphics[scale=0.4]{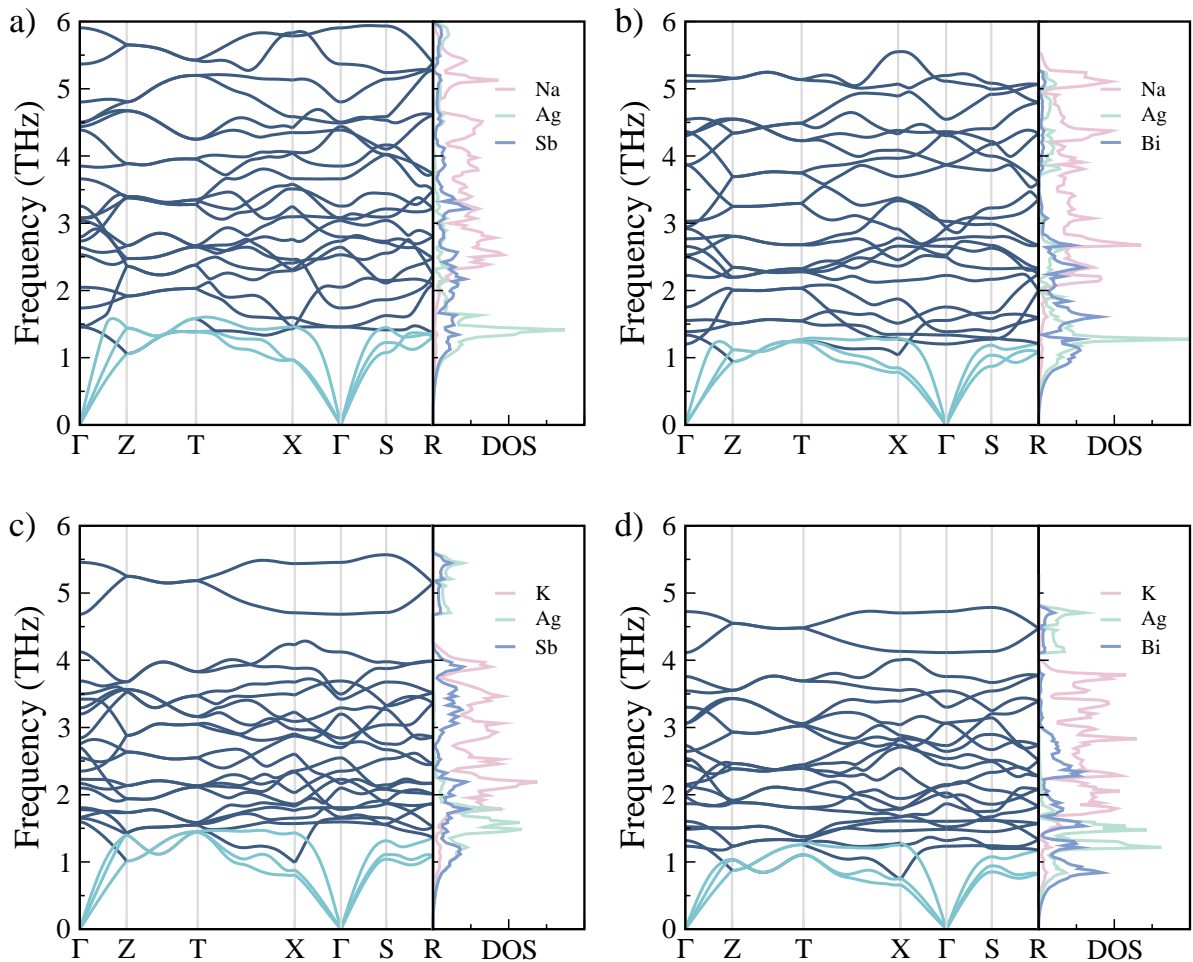}
\caption{Phonon band structures and partial density of states
of (a) Na$_2$AgSb, (b) Na$_2$AgBi, (c) K$_2$AgSb, and K$_2$AgBi.
The density of states is shown in arbitrary units.}
\label{phonons}
\end{figure*}

Na$_2$AgSb crystallizes in orthorhombic $Cmcm$ symmetry \cite{Schuster79},
while Na$_2$AgBi \cite{Wang21} is predicted to be thermodynamically stable in the
same symmetry. The K-atom based systems K$_2$AgSb and K$_2$AgBi
adopts orthorhombic $C222_1$ structure \cite{Savelsberg77}. The crystal structure in
two symmetries, $Cmcm$ and $C222_1$, bears a striking resemblance
and looks fairly identical. The only notable
distinction is that the atomic positions of Na/K and Ag atoms
are marginally off-centered along \textit{c}- and \textit{a}-axis,
respectively, as can be noted from structural parameters in Supplemental Material
\cite{Supp}. Thus, the crystal
structure of all four systems can be conveniently represented by
the structure shown in Fig.~\ref{crystal}(a). The structure can be visualized
as the infinite zigzag chains of Ag-Sb/Bi well separated from
parallel chains by Na/K atoms along \textit{a}-axis. Nevertheless, to
ensure the accuracy of our simulations, we optimized the four
systems in their respective symmetries. Our calculated lattice
parameters are in agreement with experimental values ($\pm$2\%),
as evident from Table I. Using these optimized parameters, we
calculated the electronic structure, as discussed next.

\begin{table}[]
\caption{Calculated lattice constants and band gaps. The experimental
values provided in parentheses for Na$_2$AgSb and K$_2$AgSb/Bi
are from the Ref.~\cite{Schuster79} and \cite{Savelsberg77},
respectively.}
\centering
\begin{tabular*}{\columnwidth}{l @{\extracolsep{\fill}} lllll}
\hline
\hline
System		&a (\AA)	&b (\AA)	& c (\AA)	&E$_g$ (eV) \\
\hline
Na$_2$AgSb      &9.34 (9.29)	&7.90 (7.90)	&5.82 (5.77)    &1.36	 \\
Na$_2$AgBi      &9.51           &8.04           &5.87		&0.87	\\
K$_2$AgSb       &10.61 (10.44)	&8.36 (8.29)	&6.36 (6.28)	&1.87	\\
K$_2$AgBi       &10.79 (10.56)  &8.48 (8.38)  	&6.46 (6.33)	&0.99     \\
\hline
\hline
\end{tabular*}
\end{table}

\subsection{Electronic Structure}

Understanding electronic structure is quintessential
for knowing the thermoelectric performance of a material.
The electronic structure calculated using standard DFT functionals
like PBE is known to underestimate the band gap. Therefore, we
have employed mBJ potential in our simulations. The mBJ is known
to provide band gaps with accuracy similar to hybrid functionals;
however, it is less computationally expensive. In addition, to account
for the heavier atoms, we have also incorporated spin-orbit coupling
in our simulations. Our calculated electronic
structures of four systems are depicted in Fig.~\ref{crystal}. All materials are
semiconducting in nature, bearing band gap in the range (0.8--1.8 eV).
The values seem sufficient to avoid any bipolar conduction, which
could be beneficial for thermoelectric materials. The nature
of band gap is direct for Na$_2$AgSb, Na$_2$AgBi, and K$_2$AgSb,
where valence band maximum (VBM) and conduction band minimum (CBM)
lies along $\Gamma$-point. For K$_2$AgBi, the VBM is off-gamma at
(-0.381135  0.488430  0.43478) \textit{k}-point along T-X direction,
whereas CBM lies along $\Gamma$.

The band curvature and slope provide insight into the
behavior of $S$ and $\sigma$. The steep dispersed bands
have lighter effective mass of charge carriers, which
contributes to electrical conductivity, whereas the
flat bands have higher effective mass beneficial for
the Seebeck coefficient. All the systems have more
dispersed bands in the conduction band region, which
suggests high $\sigma$ for \textit{n}-type systems.
On the other hand, the valence band region displays
a mix of mostly flat bands close to the Fermi level, followed
by some dispersed bands. Thus, \textit{p}-type systems
are likely to have high Seebeck coefficient.
The presenece of flat bands along with dispersed
ones also hints at larger power factor for \textit{p}-type
systems.


The density of states show the dominant contribution of
Sb/Bi in valence band maxima in all the systems, whereas the
conduction band minima is populated by the zigzag alignment of
Ag-Sb/Bi in Na$_2$AgSb/Bi. The conduction band minima in
K-based compositions, i.e., K$_2$AgSb/Bi, is almost equally
contributed by all the atoms.

To gain insight into bonding characteristics, we further performed
COHP analysis, Fig.~\ref{cohp}. 
This helps in predicting the bonding nature by allocating the electronic
energy into individual orbital interaction pairs. The positive and negative
COHP values represent the bonding and antibonding states, respectively.
No antibonding states are observed below the Fermi level for the plausible
interactions (Na/K-Ag, Na/K-Sb/Bi, and Ag-Sb/Bi). Across the materials, the
interactions between alkali and silver atoms (Na/K-Ag) are negligible. The interactions
between Ag-Sb/Bi are dominant, followed by Na/K-Sb interactions. This fits well with
the classical description of Zintl phases \cite{Nesper14}, where Na/K atoms are loosely bounded
to the Ag-Sb/Bi framework. The interactions between the Ag-Sb/Bi and Na-Sb/Bi
are pretty similar in Na$_2$AgSb/Bi. On the other hand, the interactions between
the Ag-Sb/Bi are more pronounced in comparison to K-Sb/Bi in K-based materials.
This reasonable disparity suggests the more bonding heterogeneity in K$_2$AgSb/Bi,
detrimental to lattice thermal conductivity.

\subsection{Phonons}

Now, we are moving on to phonon dispersion curves,
which play a pivotal role in transport properties.
The phonon dispersion curves and density of states are shown in
Fig.~\ref{phonons}. The primitive crystal structure has 8 atoms,
resulting in 24 phonon bands at each wave vector. The dynamical
stability of a crystal structure is crucial for accurately predicting
its properties. No imaginary frequencies in the phonon
dispersions suggest that the crystal structure can sustain
lattice vibrations and is unlikely to undergo a phase transition
or structural distortions. The phonon dispersions range $\sim$5.9~THz
in Na$_2$AgSb, which flattens to $\sim$4.8~THz in K$_2$AgBi. Such low
frequencies of phonon modes are likely to have low group velocities,
$\nu_g$ = d$\omega$/d$q$, catering to low $\kappa_L$. The steep acoustic
modes flatten out between $\sim$1-1.5~THz and thereafter followed by
relatively flat low-lying optical modes. It can also be seen that one of
the optical modes coincides with acoustic modes. 

It is interesting to note the band gap in high frequency optical modes
in K$_2$AgSb and K$_2$AgBi at $\sim$4.3 and $\sim$4.0~THz, respectively.
Such band gaps are associated with the flattening of phonon modes, resulting
in low group velocities and thereby low $\kappa_L$ \cite{Rodriguez23}.
Further, there is an avoided crossing between acoustic and optical modes
along the $\Gamma$-S-R direction ($\sim$1.1 THz) in all the systems except
Na$_2$AgSb. This avoided crossing is characteristic of rattling behavior
in the system. Such avoided crossings suppress the dispersion of acoustic
modes, leading to reduced slopes and lower phonon group velocities, conducive
to low $\kappa_L$ \cite{Lin16}.  

The phonon density of states help get an insight into the atomic contribution
towards various phonon modes. The acoustic modes primarily responsible for the
heat conduction in the lattice are contributed mainly by Ag in Na$_2$AgSb, with some
contribution from Sb atoms. In the case of K$_2$AgSb, the Ag and Sb contribute
almost equally to the acoustic modes, whereas Bi dominates the acoustic region
in Na$_2$AgBi and K$_2$AgBi. The mid-frequency phonons see a dominant contribution
from the alkali atoms Na/K, whereas the high frequency optical modes are contributed
by the lighter Na atoms in Na$_2$AgSb and Na$_2$AgBi. Surprisingly, the high frequency
optical modes above the band gap are contributed by Ag and Sb/Bi in K-based systems
K$_2$AgSb and K$_2$AgBi despite the lighter mass of K atoms, which could be a subject
of further study. Next, we discuss the trend of $\kappa_L$ for different systems.

\begin{figure}
\centering\includegraphics[scale=0.45]{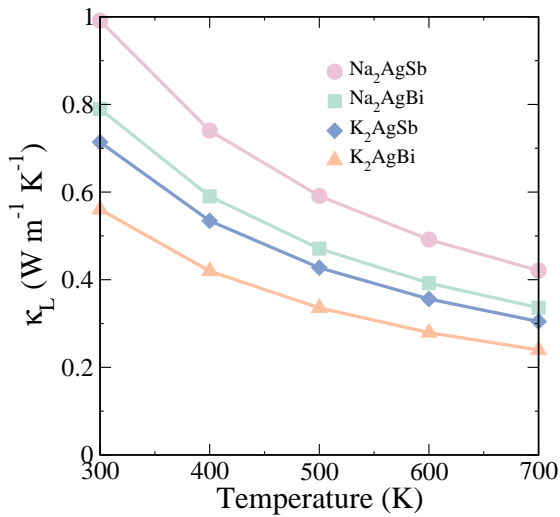}
\caption{Average lattice thermal conductivity as a function
of temperature for Na$_2$AgSb, Na$_2$AgBi, K$_2$AgSb, and K$_2$AgBi.}
\label{kappa}
\end{figure}

\begin{figure}
\centering\includegraphics[scale=0.35]{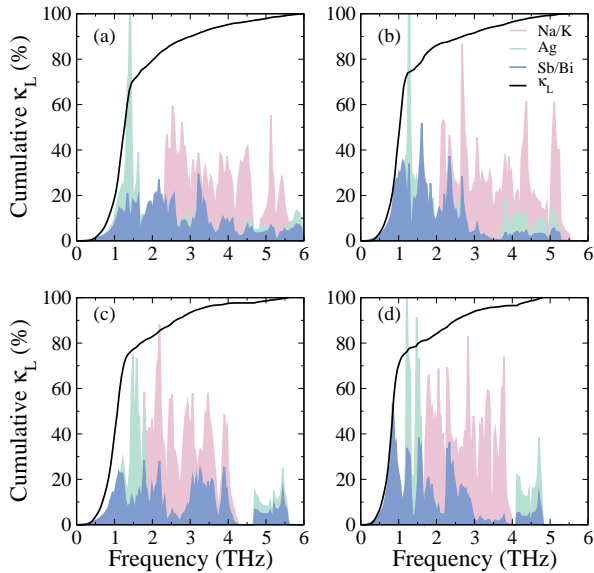}
\caption{Cumulative lattice thermal conductivity in \% as a function
of phonon frequency at 300~K for (a) Na$_2$AgSb, (b) Na$_2$AgBi, (c) K$_2$AgSb,
and (d) K$_2$AgBi. The phonon density of states in arbitrary units are overlaid.}
\label{kc}
\end{figure}

\begin{figure*}
\centering\includegraphics[scale=0.6]{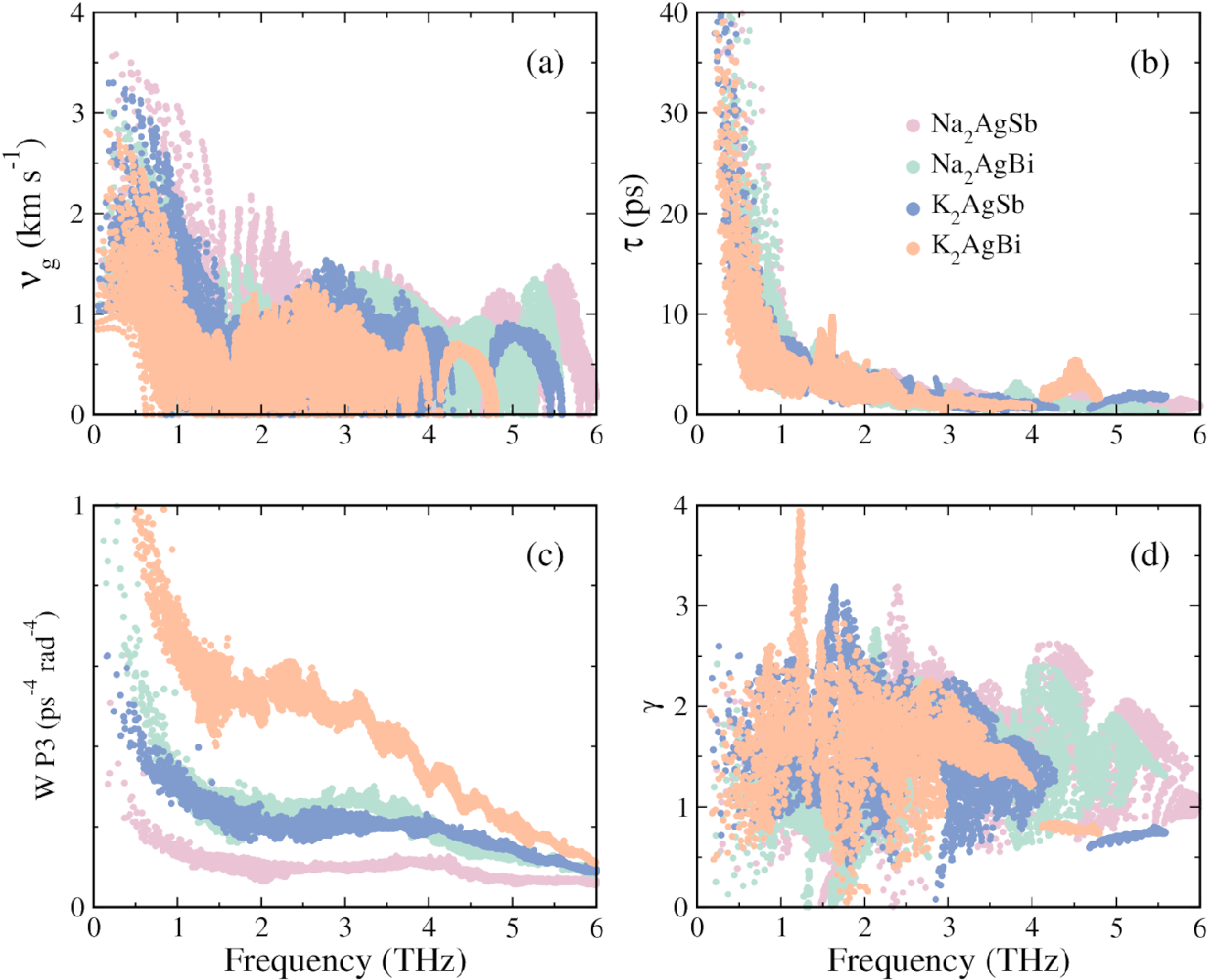}
\caption{(a) Phonon group velocities, (b) lifetime, (c) weighted scattering
phase space, and (d) Gr{\"u}neisen parameter as a function of frequency for
Na$_2$AgSb, Na$_2$AgBi, K$_2$AgSb, and K$_2$AgBi.}
\label{gv}
\end{figure*}

\subsection{Lattice Thermal Conductivity}

Figure~\ref{kappa} illustrates the trend of average $\kappa_L$ as a function
of temperature for different systems. The average values are obtained using
the arithmetic mean along different crystallographic directions. Our calculated
values can be seen as an upper bound, considering the RTA method takes into account
only the three-phonon scattering processes. It has been demonstrated that the
four-phonon scattering process further reduces the $\kappa_L$ \cite{Feng17, Xia18, Yue23}.
The $\kappa_L$ values are also subject to the real material, which may have defects, grain
boundaries, and other imperfections. All these said factors are likely to lower
the $\kappa_L$ further.

For all the systems, a decreasing trend of $\kappa_L$ with temperature can be
noticed. The behavior is consistent with the more pronounced phonon-phonon
scattering at higher temperatures. The lower values at high temperature
signify the efficient thermoelectric performance at high temperatures. Across
the materials, Na$_2$AgSb exhibit the highest $\kappa_L$ values starting at
0.9 W m$^{-1}$ K$^{-1}$ at 300~K and decreasing to 0.4 W m$^{-1}$ K$^{-1}$
at 700~K. The Na$_2$AgBi and K$_2$AgSb show intermediate $\kappa_L$ values
both starting at around 0.7 W m$^{-1}$ K$^{-1}$ at 300~K and falling to
approximately 0.3 W m$^{-1}$ K$^{-1}$ at 700~K. The standout material is
K$_2$AgBi exhibiting strikingly low $\kappa_L$ values ranging 0.56--0.23
W m$^{-1}$ K$^{-1}$ in the temperature window 300--700~K.  
Our measured $\kappa_L$ values are significantly low and compete well
with established thermoelectric materials SnSe \cite{Zhao14} ($\kappa_L$ $\sim$ 0.47
W m$^{-1}$ K$^{-1}$ at 300~K), PbTe \cite{Bessas12} ($\kappa_L$ $\sim$ 2.0 W m$^{-1}$ K$^{-1}$
at 300~K), Bi$_2$Te$_3$ \cite{Wang11} ($\kappa_L$ $\sim$ 1.6 W m$^{-1}$ K$^{-1}$ at 300~K),
and also with other low $\kappa_L$ reported Zintl phases \cite{Toberer10, Zevalkink13, Zheng22}. 


To gain insight into such astonishingly low $\kappa_L$ values, we first
understand the individual atomic and phonon modes contribution to $\kappa_L$.
This helps optimize the material's performance by possibly reducing the
$\kappa_L$ through desirable structural modifications. As discernible from
Fig.~\ref{kc}, the cumulative nature of the curves shows that approximately
80\% contribution to $\kappa_L$ comes from acoustic and low-lying optical modes
below 2 THz. Notably, the acoustic modes alone contribute $\sim$71 to 77\%
towards $\kappa_L$ on going from Na$_2$AgSb to K$_2$AgBi. There is a marked
decrease in contributions from high frequency optical modes. Now, these primary
heat carrying acoustic modes are contributed majorly by Ag in Na$_2$AgSb, Ag
and Bi in Na$_2$AgBi, Ag and Sb in K$_2$AgSb, and Bi in K$_2$AgBi. Thus, any
attempt to further reduce $\kappa_L$ should aim to alloy at these respected
sites.

Now, we move to another critical parameter in understanding the $\kappa_L$,
i.e., phonon group velocity, $\nu_g$. The group velocity typically decreases
with increasing phonon frequency, as discernible from Fig.~\ref{gv}(a). This could
be attributed to the more scattering tendency of higher frequency phonons, which
hinders energy transport. It can also be observed that group velocity decreases
with increasing atomic mass on going from Na$_2$AgSb to K$_2$AgBi. This is because
the heavier atoms are likely to undergo smaller vibrations. The low group velocities
in K$_2$AgSb and K$_2$AgBi, as discussed earlier, can also be attributed to the gap in
phonon modes, associated with the flattening of phonon modes and resulting in low group
velocities. The majority of the acoustic modes (up to $\sim$1.6 THz) possess low group
velocities ($\nu_g<$ 1.5 km s$^{-1}$). As already seen, the acoustic modes contributing
most to the $\kappa_L$, the low group velocities of acoustic modes could be valuable.
On the other hand, the bulk mid-frequency phonons exhibit $\nu_g<$~1 km s$^{-1}$,
whereas the most high frequency optical modes exhibit even lesser $\nu_g$.

The sound velocities obtained from the slopes of the acoustic modes allow a
better comparative analysis of our results with other reported works. The
average values are obtained using the expression
$\frac{3}{\nu_s^3} = \frac{1}{\nu_l^3} + \frac{2}{\nu_t^3}$,
where $\nu_l$ and $\nu_t$ are the longitudinal and transverse sound
velocities, respectively. Our
calculated sound velocities for Na$_2$AgSb, Na$_2$AgBi, K$_2$AgSb, and
K$_2$AgBi are 2.2, 1.9, 2.0, and 1.7 km s$^{-1}$, respectively. Such
remarkably low sound velocities are comparable to standard thermoelectric
materials such as SnSe \cite{Zhao14} (3.1 km s$^{-1}$), PbTe \cite{Bessas12}
(1.7 km s$^{-1}$), and Bi$_2$Te$_3$ \cite{Wang11} (1.5 km s$^{-1}$).

It is instructive to examine the phonon lifetimes as they are linearly
proportional to the $\kappa_L$ by Eqn.~\ref{eqn}. The phonon lifetimes
as extracted using three-phonon scattering processes are depicted in
Fig.~\ref{gv}(b). The lifetime of acoustic modes ($\sim$1-1.5 THz) is nearly
one order higher than the optical modes, indicative of the dominant contribution
of acoustic modes to $\kappa_L$. The high frequency optical phonon modes ($>$3~THz)
tend to scatter readily due to interactions such as Umklapp processes, which result
in smaller lifetimes. Evidently, the lifetime of Na$_2$AgSb and Na$_2$AgBi across the
spectrum are higher in comparison to K$_2$AgSb and K$_2$AgBi, partially accounting for
their higher $\kappa_L$. It is obvious to note that Na$_2$AgBi and K$_2$AgBi exhibit lower
frequencies in the optical phonon branches compared to their Sb counterparts, suggesting
increased phonon-phonon scattering. Further, the K-based systems (K$_2$AgSb and K$_2$AgBi)
shows lower phonon lifetimes than their Na-based counterparts. Thus, it can be surmised that
substituting Na/Sb with K/Bi may reduce the phonon lifetime and $\kappa_L$.
The majority of phonon modes have lifetimes of less than 10 ps, which suggests anharmonicity
and strong phonon-phonon scatterings. The average lifetimes across the systems are less than
$\sim$3~ps, fairly consistent with SnSe \cite{Guo15} (3.2~ps at 300~K).

To delve deeper into understanding the trend of $\kappa_L$, we next investigate the
phase space, which describes the available states for phonon scattering, weighted
according to interaction probabilities. A large phase space is an indication of
more phonon scattering events. As discernible from Fig.~\ref{gv}(c), K$_2$AgBi has the
largest phase space among the materials. Indeed, the phase space for K$_2$AgBi is nearly
double the size of Na$_2$AgBi and K$_2$AgSb, and almost four times that of Na$_2$AgSb.
Verily, this accounts for the lowest $\kappa_L$ in K$_2$AgBi. The phase space trend
also highlights how substituting heavier atom (Na/Sb by K/Bi) influences scattering
behavior. The large phase space of Na$_2$AgBi (K$_2$AgBi) in comparison to Na$_2$AgSb
(K$_2$AgSb) suggests that heavier atoms shift phase space distribution. 
This very fact can be leveraged in material design to tune the desirable properties
through phonon scattering control. The large phase space values also indicate the strong
anharmonicity in the systems.

To quantify the same, we evaluated the Gr{\"u}neisen parameter ($\gamma$), which is a measure of
anharmonicity in the crystal lattice. As discernible from Fig.~\ref{gv}(d), the majority of phonon
modes have $\gamma$ greater than unity. The average values of $\gamma$ are 1.53, 1.40, 1.48, and 1.57
for Na$_2$AgSb, Na$_2$AgBi, K$_2$AgSb, and K$_2$AgBi, respectively. These values align closely
with those reported for materials exhibiting low $\kappa_L$, where significant anharmonicity
dominates the lattice vibrations. For instance, SnSe ($\gamma\sim$1.7), Bi$_2$Te$_3$ ($\gamma\sim$1.5),
and PbTe ($\gamma\sim$1.4) exhibit comparable values, highlighting the strong anharmonicity in these
systems. Meanwhile, the largest $\gamma$ for K$_2$AgBi can be attributed to large values of $\gamma$
in the acoustic region. The $\gamma$ in the acoustic region reaches $\sim$4 for K$_2$AgBi, whereas other
materials have a maximum $\gamma<$ 2.6. The large $\gamma$ of K$_2$AgBi also justifies its low $\kappa_L$
in comparison to other materials.

The promising thermal transport warrants further assessment of electrical transport
properties to obtain thermoelectric efficiency. In the following section, we
evaluate and understand the electrical transport coefficients.

\subsection{Electrical Transport Properties}

\begin{figure*}
\centering\includegraphics[scale=0.4]{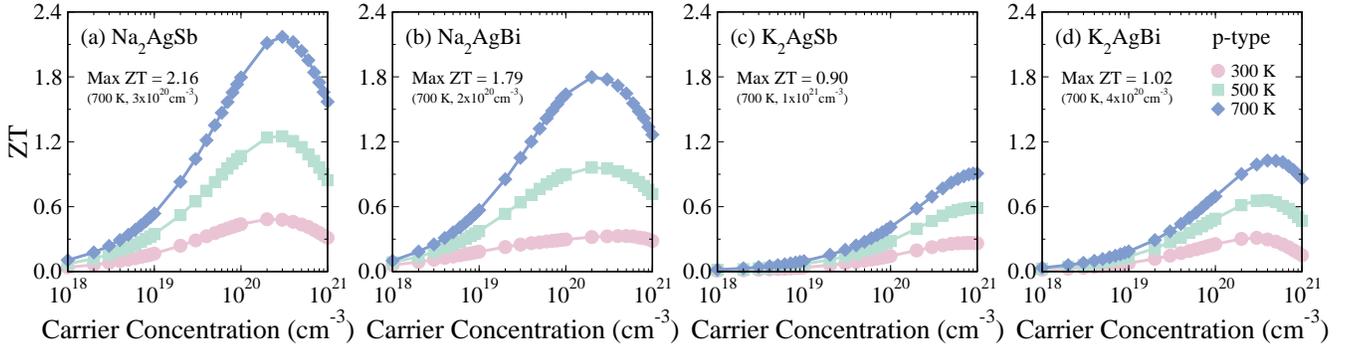}
\caption{Thermoelectric figure of merit as a function of temperature at various carrier concentrations of \textit{p}-type (a) Na$_2$AgSb, (b) Na$_2$AgBi, (c) K$_2$AgSb, and (d) K$_2$AgBi.}
\label{zt}
\end{figure*}

In this section, we will discuss the temperature and carrier-concentration dependent
electrical transport properties obtained by solving the Boltzmann transport equation
using inputs from density functional theory. Given the fact that Zintl phases have a
natural tendency to form \textit{p}-type compositions \cite{Brown06, Wang07}
and a glimpse of better
\textit{p}-type transport properties from the electronic structure discussion, we
emphasize here our results for \textit{p}-type systems. Interested readers are directed
to Supplemental Material \cite{Supp} for the \textit{n}-type systems. We take the liberty of discussing
our results at 300, 500, and as high as 700 K, based on the synthesis temperature of 973~K
(Na$_2$AgSb) and 873~K (K$_2$AgSb and K$_2$AgBi). Figure~\ref{zt} shows our calculated
results for the figure of merit at different temperatures and carrier concentrations.

As discernible from the figure, the thermoelectric figure of merit increases with
temperature for the given carrier concentration range of 10$^{18}$ to 10$^{21}$ cm$^{-3}$.
Our predicted maximum values of the figure
of merit are, $ZT\sim$2.1, 1.7, 0.9, and 1.0 at 700~K for Na$_2$AgSb, Na$_2$AgBi, K$_2$AgSb, and
K$_2$AgBi, respectively, for carrier concentration in the range of 10$^{20}$ cm$^{-3}$.
The comparatively lower values for K$_2$AgSb and K$_2$AgBi highlight the influence of
Na substitution with K on thermoelectric performance. Nevertheless, these
values are substantially high and compete well with contemporary Zintl phases, e.g.,
Mg$_3$Sb$_2$ \cite{Tamaki16} ($ZT\sim1.5$ at 716~K), Yb$_{10}$MgSb$_9$ \cite{Borgs23}
($ZT\sim1.0$ at 873~K), NaCdSb \cite{Guo23} ($ZT\sim1.3$ at 673~K), and BaCuSb \cite{Zheng22}
($ZT\sim0.4$ at 1010~K). Besides, the predicted values also resonate well
with other state-of-the-art materials such as half-Heusler alloys ($ZT\sim$1.4 and 1.5 at 973~K for
ZrCoBi \cite{Zhu18} and TaFeSb \cite{Zhu19}, respectively), skutterudite CoSb$_3$ \cite{Pang24}
($ZT\sim$1.7 at 823~K), PbTe-SrTe \cite{Tan16} ($ZT\sim$2.5 at 923~K), and SnSe \cite{Zhao14}
($ZT\sim$2.6 at 913~K).

\begin{figure}
\centering\includegraphics[scale=0.35]{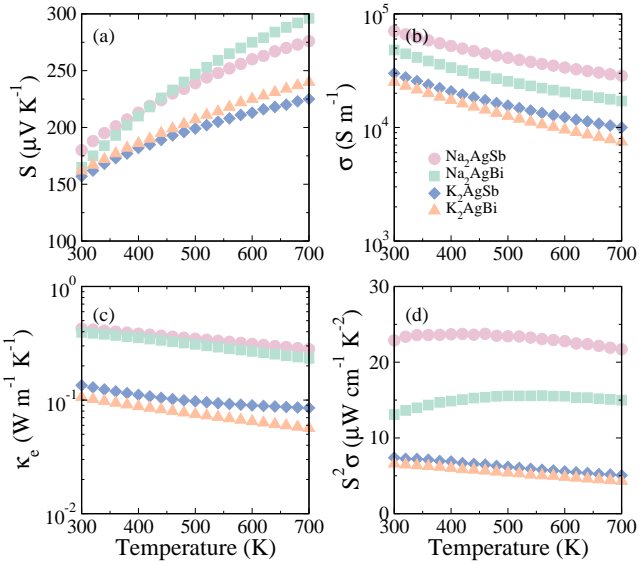}
\caption{(a) Seebeck coefficient, (b) electrical conductivity, (c) electronic thermal conductivity,
and (d) power factor as a function of temperature at optimal carrier concentrations for Na$_2$AgSb
(\textit{n} = 3$\times$10$^{20}$ cm$^{-3}$), Na$_2$AgBi (2$\times$10$^{20}$ cm$^{-3}$),
K$_2$AgSb (1$\times$10$^{21}$ cm$^{-3}$), and K$_2$AgBi (4$\times$10$^{20}$ cm$^{-3}$).}
\label{pf}
\end{figure}

Having said that, the optimal carrier concentrations are slightly on the higher side for
a doped semiconductor. Nonetheless, considering the experimental challenges, one can target
slightly lower carrier concentrations ($\sim10^{19}$ cm$^{-3}$) to breach the figure of
merit of unity for Na-based compounds.
It is noteworthy that despite having the lowest $\kappa_L$, the $ZT$ of K$_2$AgBi is also the lowest.
To dissect this, we now break down into individual components of electrical transport.
The coefficients $S$, $\sigma$, $\kappa_e$, and $S^2\sigma$ for the optimal carrier concentrations
with respect to temperature are shown in Fig.~\ref{pf}. The Seebeck coefficient represents the
change in voltage with respect to temperature and is generally expressed as
\begin{equation}
	S = \frac{8\pi^2 k_B^2}{3eh^2} m^*_{DOS} T \left(\frac{\pi}{3n}\right)^{2/3}
\label{S}
\end{equation}
where $k_B$, $e$, $h$, $n$, $m^*_{DOS}$, represents the Boltzmann constant, elementary charge,
Planck’s constant, carrier concentration, and effective mass of charge carrier, respectively.
The Eq.~\ref{S} shows the contradictory dependence of $S$ on carrier concentration and
temperature. We have observed a similar trend for the calculated values.
As discernible from Fig.~\ref{pf}(a), the $S$ increases
with temperature for all the systems. Further, beyond 300~K, the $S$ gradually
picks up for Na$_2$AgSb/Bi in comparison to K$_2$AgSb/Bi on account of slightly lower optimal
carrier concentrations.
Na$_2$AgBi exhibits the highest value ($\sim$296 $\mu$V K$^{-1}$), closely followed by Na$_2$AgSb.

The electrical conductivity reflects the ease of carrier transport in a material and is expressed
as $\sigma$ = $ne\mu$, where $n$ is the carrier concentration, $e$ is the elementary charge, and
$\mu$ is the carrier mobility. Na$_2$AgSb and Na$_2$AgBi exhibit the highest conductivity values,
with Na$_2$AgSb slightly outperforming Na$_2$AgBi, Fig.~\ref{pf}(b). Surprisingly, the $\sigma$
values for K$_2$AgSb/Bi are approximately half in comparison to Na$_2$AgSb/Bi.
The $\sigma$ of a semiconductor is known to increase with temperature and carrier concentration.
However, for degenerate semiconductors, the $\sigma$ decreases with temperature, as observed
in the present case. This decreasing behavior can be attributed to increased carrier scattering at
higher temperatures, which impacts the carrier mobility, thereby $\sigma$.

The electronic thermal conductivity, $\kappa_e$, accounts for the contribution of charge carriers
to heat transport and are consistent with the electrical conductivity trend, Fig.~\ref{pf}(c).
$\kappa_e$ decreases with temperature, likely due to increased scattering of carriers. Na$_2$AgSb
and Na$_2$AgBi have higher values of $\kappa_e$. In conjunction with Fig.~\ref{kappa}, it can be
stated that heat propagation through lattice is the dominating mechanism to total thermal conductivity,
a typical behavior of semiconductors.

Now coming to the power factor ($S^2\sigma$), which serves
as a critical metric for thermoelectric performance, accounting
for both $S$ and $\sigma$. As discernible from Fig.~\ref{pf}(d),
the power factor begins to plateau near 400~K (450~K) for
Na$_2$AgSb (Bi), followed by a decrease at higher temperature.
However, the power factor decreases across the temperature range
for K$_2$AgSb/Bi. We obtained highest power factor for Na$_2$AgSb
across the temperature range, peaking near 23 $\mu$W K$^{-1}$ cm$^{-2}$
at 420~K. Na$_2$AgBi follows Na$_2$AgSb, indicating their superior
potential for thermoelectric applications, whereas the power factor
for K-based materials (K$_2$AgSb and K$_2$AgBi) are substantially low
in comparison to Na$_2$AgSb/Bi. The possible reason for such a trend
could be the dramatic decline in $\sigma$ values, as noted from
Fig.~\ref{pf}(b). To reason out such low $\sigma$ values in K$_2$AgSb/Bi,
we discuss next the scattering rates and mobility of charge carriers for
the optimal carrier concentrations. 

\begin{figure}
\centering\includegraphics[scale=0.35]{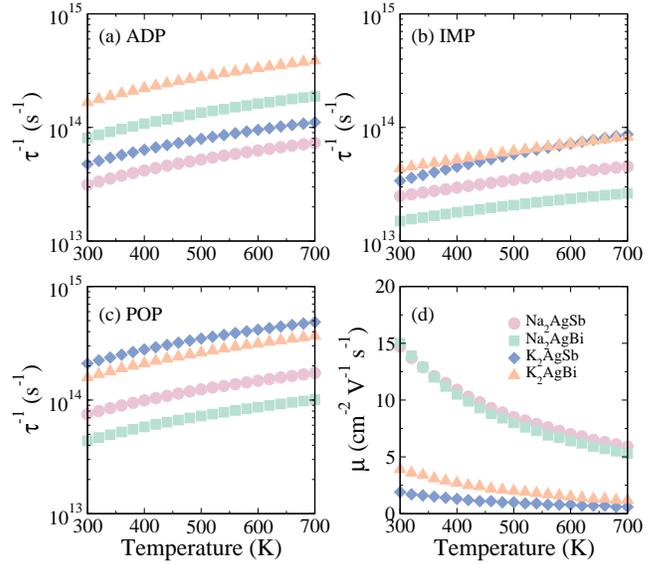}
\caption{(a), (b), (c) represents the scattering rates for ADP, IMP, POP scattering mechanisms,
respectively, and (d) mobility of charge carriers as a function of temperature for optimal carrier
concentrations, i.e., Na$_2$AgSb
(\textit{n} = 3$\times$10$^{20}$ cm$^{-3}$), Na$_2$AgBi (2$\times$10$^{20}$ cm$^{-3}$), 
K$_2$AgSb (1$\times$10$^{21}$ cm$^{-3}$), and K$_2$AgBi (4$\times$10$^{20}$ cm$^{-3}$).}
\label{scat}
\end{figure}

We include acoustic deformation potential (ADP), ionized impurity (IMP), and
polar optical phonon (POP) scattering mechanisms to capture the effect of lattice vibrations
on charge carriers, the impact of charged impurities on carrier mobility, and interactions
between charge carriers and the electric field generated by polar optical phonons, respectively. 
Fig.~\ref{scat} illustrates our results of the scattering rates and carrier mobility as a function
of temperature for the optimal carrier concentrations. The scattering rates for all mechanisms
(ADP, IMP, and POP) increase steadily with temperature, suggesting enhanced interactions of
charge carriers with phonons at higher thermal energies. The ADP is the most dominating
mechanism for Na$_2$AgBi and K$_2$AgBi, whereas the POP mechanism dominates in the case of Na$_2$AgSb
and K$_2$AgSb. This suggests that the scattering of charge carriers by acoustic phonons is more prominent
for Bi-based systems, while charge carriers in Sb-based systems are more prone to scattering through
optical phonons. On the other hand, the IMP scattering is least affected by temperature.
Overall, K$_2$AgSb and K$_2$AgBi depict higher scattering rates in comparison to Na-based counterparts.
The mobility trends in Fig.~\ref{scat}(d) reveal a significant reduction with increasing temperature
for all systems, driven by the cumulative effect of rising scattering rates. It should be noted that
in addition to scattering rates, the mobility of carriers is also dependent on band dispersion and
effective mass. Nonetheless, Na$_2$AgSb and Na$_2$AgBi maintain comparatively higher mobilities,
underscoring their superior charge transport properties under the examined conditions.

The mobility trend mirrors the $\sigma$ pattern (Fig.~\ref{pf}(b)) and also explains
lower $\sigma$ and power factor values of K$_2$AgSb and K$_2$AgBi. Such low mobilities
of K-based materials can also be traced to the band dispersions. The bands close to
the Fermi level in the valence band region are relatively flat for K$_2$AgSb/Bi in comparison
to Na$_2$AgSb/Bi, Fig.~\ref{crystal}. This analysis highlights the critical influence
of scattering mechanisms and carrier concentration optimization on the transport
properties of proposed thermoelectric materials, offering valuable insights for
achieving high-performance thermoelectric efficiency.

\section{Discussion}

Our theoretical predictions suggest exceptionally good high temperature thermoelectric
performance for the investigated materials, especially Na$_2$AgSb and Na$_2$AgBi. However,
the theoretical prediction of exotic properties should meet the experimental realization, which is
subject to various challenges. As these materials (except Na$_2$AgBi) are already synthesized,
the material fabrication should not be deemed a major concern. However, achieving the proposed
optimal carrier concentrations could be the prime obstacle in meeting the desired figure of merit.
Two prominent ways of controlling the carrier concentration are extrinsic doping or tuning the
intrinsic defects. Yb$_{14}$Mn$_{1-x}$Al$_x$Sb$_{11}$ \cite{Cox09} and
Ca$_x$Yb$_{1-x}$Zn$_2$Sb$_2$ \cite{Gascoin05} are examples
of some well-known extrinsic \textit{p}-type Zintl phases, offering the feasibility of creating
the extrinsic compositions. In light of such evidence, group-IV elements (like Si, Ge, Sn) in place
of Sb/Bi or group-III elements (e.g., Al, Ga) in place of Ag could be the most favorable choices
for achieving extrinsic \textit{p}-type materials. On the other hand, the Zintl phases are also
found to have intrinsic \textit{p}-type compositions. For instance, Toberer \textit{et al}.
\cite{Toberer09}
synthesized \textit{p}-type LiZnSb samples, with a carrier concentration of the order of
10$^{20}$ carriers cm$^{-3}$. In the case of Na$_2$AgSb/Bi and K$_2$AgSb/Bi, the loss of Na/K
ions to oxidation during synthesis could result in intrinsic \textit{p}-type materials. The absence
of four Na/K ions will correspond to 0.04 carriers per formula unit, which is approximately
of the order of 10$^{20}$ carriers cm$^{-3}$. Such doping levels are also equivalent to our
proposed carrier concentrations, suggesting their experimental realization.  
In addition, the credibility of computational methods used for predicting materials' properties
is also of key importance.

In this study, we have used state-of-the-art computational techniques
which are well tested with experimental works for a wide range of materials
\cite{Togo15, Skelton14, Rundle22, Flitcroft22, Dou24}.
However, we have not
taken into account the anharmonic phonon renormalization due to
computationally expensive requirements. We are of the
opinion that it will be important to qualitatively discuss the impact of anharmonic phonon
renormalization on our predicted results. It has been found in various studies that anharmonic phonon
renormalization impacts more the low frequency phonons \cite{Yao23, Guo_SnSe}.
This suggests stronger anharmonicity in the
low frequency phonons, suggesting increased acoustic-optical phonons interactions. This could eventually
lead to four-phonon scattering mechanism, which is found to further diminish the lattice thermal
conductivity. Studies have found that four-phonon processes are more impactful at high temperatures
and frequencies \cite{Feng17, Xia18, Yue23}. Our calculated P3-phase space further suggests
three-phonon scattering mechanism
as more influential in low frequency region, i.e., less than 2~THz. As low frequency acoustic modes
are major contributors to heat conduction, it can be surmised that four-phonon processes are likely
to lower the lattice thermal conductivity at higher frequencies and temperatures. 

Nevertheless, our results show the importance of chemical composition
and optimal carrier concentrations in optimizing thermoelectric properties.
The better performance of Na-based systems suggests that light alkali metal cations,
combined with Ag-Bi/Sb frameworks, could be effective in achieving high thermoelectric
efficiency. Furthermore, the differences between Na- and K-based systems highlight the
tunability of transport properties through cation substitution. This could pave the way for material
design for specific operating conditions.
The promising performance of Na$_2$AgSb and Na$_2$AgBi warrants further experimental
validation and exploration of strategies to achieve the optimal carrier concentrations.
Additionally, the role of phonon scattering mechanisms in suppressing lattice thermal
conductivity in these compounds should be investigated using advanced techniques such
as anharmonic phonon calculations.

\section{Summary}

In conclusion, we studied the alkali metal-based family of Zintl phases X$_2$AgY
(X= Na, K; Y = Sb, Bi) using density functional theory based first-principles
simulations and by solving Boltzmann transport equations. The Na- and K-based
materials crystallize in orthorhombic $Cmcm$ and $C222_1$ symmetry, respectively.
All materials are semiconducting, possessing band gap in the range of 0.8-1.8 eV,
sufficient to avoid any bipolar conduction. The dynamical stability of the systems
is ascertained by phonon dispersions. The main highlight of the work is the remarkably
low lattice thermal conductivity values ($<$ 1 W m$^{-1}$ K$^{-1}$), comparable to
benchmark thermoelectric materials such as SnSe, PbTe, and Bi$_2$Te$_3$.
It is found that around 80\% contribution to thermal conductivity comes from acoustic
and low-lying optical modes. We ascribed low thermal conductivity values to low phonon
velocities, short lifetimes, and large scattering phase space. Notably, an avoided
crossing in the phonon spectrum, indicative of rattling behavior, was observed in all
systems except Na$_2$AgSb. The rattling phenomenon is associated with suppressing the dispersion
of acoustic modes, which could lead to low phonon velocities. In addition, in K-based materials,
a gap in optical modes is found, which often results in the flattening of phonon modes.
Furthermore, the observed bonding heterogeneity also accounts for their low
thermal conductivity. Combining
these properties with electrical transport, we determined a high figure of merit at 700~K, i.e.,
$ZT\sim$ 2.1 for Na$_2$AgSb, 1.7 for Na$_2$AgBi, 0.9 for K$_2$AgSb, and 1.0 for K$_2$AgBi.
Our study underscores the importance of chemical composition and carrier engineering in
optimizing thermoelectric materials.

\begin{acknowledgments}

M. Z. is thankful to SERB-DST (File No. PDF/2022/002559) for the
financial assistance. B. K. M. acknowledges the funding support
from the SERB, DST (ECR/2016/001454). The calculations are performed
using the High Performance Computing cluster, Padum, at the Indian
Institute of Technology Delhi.

\end{acknowledgments}

\bibliography{X2AgY}

\end{document}